\documentclass[conference]{IEEEtran}
\IEEEoverridecommandlockouts
\usepackage{cite}
\usepackage{amsmath,amssymb,amsfonts}
\usepackage{algorithmic}
\usepackage{graphicx}
\usepackage{textcomp}
\usepackage{xcolor}
\usepackage{dblfloatfix}
\usepackage{makecell}
\usepackage{booktabs}

\usepackage[caption=false,font=normalsize,labelfont=sf,textfont=sf]{subfig}
\usepackage{siunitx}
\def\BibTeX{{\rm B\kern-.05em{\sc i\kern-.025em b}\kern-.08em
    T\kern-.1667em\lower.7ex\hbox{E}\kern-.125emX}}

\begin{document}

\title{See Further Than CFAR: a Data-Driven Radar Detector Trained by Lidar}

\DeclareRobustCommand*{\IEEEauthorrefmark}[1]{\raisebox{0pt}[0pt][0pt]{\textsuperscript{\footnotesize #1}}}

\author{\IEEEauthorblockN{
Ignacio Roldan\IEEEauthorrefmark{1},   
Andras Palffy\IEEEauthorrefmark{1-2},    
Julian F. P. Kooij\IEEEauthorrefmark{2},
Dariu M. Gavrila\IEEEauthorrefmark{2},
Francesco Fioranelli\IEEEauthorrefmark{1},
Alexander Yarovoy\IEEEauthorrefmark{1}
}                                     
\IEEEauthorblockA{
$^{1}$Microwave Sensing, Signals and Systems (MS3) Group, Department of Microelectronics\\ $^{2}$Intelligent Vehicles (IV) Group, Department of Cognitive Robotics \\ Delft University of Technology, Delft, The Netherlands}
  
 \IEEEauthorblockA{ \emph{\{i.roldanmontero,
a.palffy, 
j.f.p.kooij,
d.m.gavrila,
f.fioranelli, 
a.yarovoy\}@tudelft.nl}}
}

\maketitle

\begin{abstract}

%
%
In this paper, we address the limitations of traditional constant false alarm rate (CFAR) target detectors in automotive radars, particularly in complex urban environments with multiple objects that appear as extended targets.

We propose a data-driven radar target detector exploiting a highly efficient 2D CNN backbone inspired by the computer vision domain. Our approach is distinguished by a unique cross-sensor supervision pipeline, enabling it to learn exclusively from unlabeled synchronized radar and lidar data, thus eliminating the need for costly manual object annotations.

Using a novel large-scale, real-life multi-sensor dataset recorded in various driving scenarios, we demonstrate that the proposed detector generates dense, lidar-like point clouds, achieving a lower Chamfer distance to the reference lidar point clouds than CFAR detectors. Overall, it significantly outperforms CFAR baselines detection accuracy.
\end{abstract}


\section{Introduction}
In recent years, the landscape of automotive technology has witnessed a transformative shift, with the integration of advanced sensor systems playing a pivotal role in enhancing vehicle safety and autonomy. Among these sensors, radar has emerged as a cornerstone technology, contributing significantly to the evolution of intelligent transportation systems. Once mostly confined to defence applications, radar is routinely used for autonomous driving in collision avoidance, adaptive cruise control, and overall vehicular awareness \cite{Bilik2019}. Unlike sensing modalities such as cameras and lidar, radar can operate effectively in adverse environmental and weather conditions with low visibility. This adaptability positions radar as a robust and reliable solution for real-time perception in the complex and dynamic context of urban and highway driving \cite{Sun2020,Palffy2020}.

However, despite its merits, integrating radar into automotive systems is challenging. One such challenge arises from the use of Constant False Alarm Rate (CFAR) detectors to generate the radar point cloud from the dense radar cube. CFAR detectors are optimal in specific scenarios \cite{Richards2015} but have limitations in the context of automotive radar \cite{Yoon2019}. 
CFAR algorithms, designed to maintain a constant false alarm rate in the presence of varying clutter conditions, may not be well-suited for the dynamic and rapidly changing environments encountered on roadways. Issues such as non-homogeneous clutter, target masking, and shadowing can compromise the efficacy of CFAR in automotive radar applications. Moreover, CFAR detectors have the inherent problem of a fixed, predefined, expected target size, given how the guard and training cells are defined. In the automotive context, the targets of interest have very different sizes, from small targets such as pedestrians, to large trucks. On top of that, the perceived size in the angular dimension depends on the range, and objects that occupy many cells in the near field may be point-like targets in the farther area. 

\begin{figure*}[!t]
\centerline{\includegraphics[width=\linewidth]{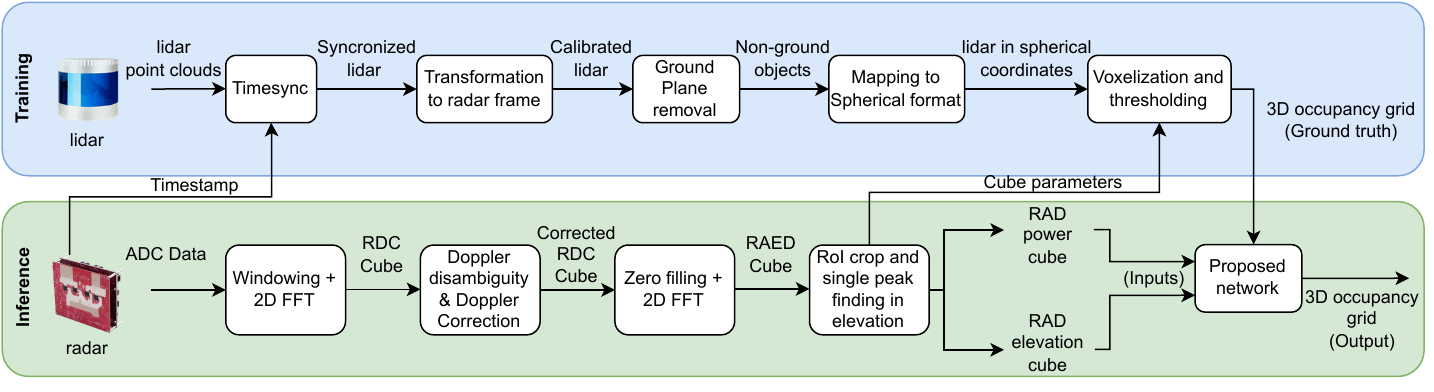}}
\caption{Overview of the proposed method. The steps to generate the 3D lidar occupancy grid are on the top row, which will be used as ground truth for training the neural network. The radar signal processing pipeline is shown at the bottom, and is needed to generate the input data. RDC stands for range-Doppler-channel (no angle estimation), and RAED stands for range-azimuth-elevation-Doppler.}
\label{High-Level-Block}
\end{figure*}

To overcome the limitations of CFAR detectors, a novel data-driven detection approach based on deep learning has been designed and tested in this work using real automotive radar data.
Data-driven detectors have been previously explored in the open literature \cite{Brodeski2019,Cheng2022,Lin2023}; however, this work introduces some new concepts that, to the best of the authors' knowledge, are investigated for the first time, namely:

\begin{itemize}
  \item The use of the full 4D range-azimuth-elevation-Doppler space as input, ensuring that no weak targets are omitted due to an initial coarse thresholding.
  \item The usage of pre-processed lidar data as ground truth to enhance radar detection.
  \item The comparison against conventional CFAR detectors using grid-based probabilities of detection and false alarm, as well as point-cloud level metrics that capture the spatial similarities between lidar and radar data.
  \item The verification with a large-scale multi-sensor dataset, including realistic automotive scenarios. This dataset, containing lidar, radar, GPS, and camera data will be published for the research community upon publication.
\end{itemize}
The rest of the paper is organized as follows. Section  \ref{sec:method} describes the proposed data-driven detector. Section \ref{sec:data} presents the data collected for the validation of the proposed approach, with results presented in Section \ref{sec:results}. Finally, Section \ref{sec:conclusion} concludes the paper.

\section{Proposed Method}
\label{sec:method}

The radar detection problem can be formulated as a binary decision task per cell in the radar cube. Before the proposed method is introduced, some definitions are needed to clarify the terminology used in the rest of the paper:

\begin{itemize}
  \item \textit{Radar cube} refers to the spherical, voxelized representation of the radar data, meaning the range, azimuth, elevation, and Doppler estimation have already been performed. Each voxel in the radar cube contains information about the reflected power in that cell.
  \item \textit{3D occupancy grid} refers to a binary cube, also in spherical coordinates, which contains ones in voxels where there are targets and zeros otherwise. This can be generated with lidar data serving as ground truth or with radar data being the main task of this work.
  \item \textit{Point cloud} refers to the Cartesian coordinates set of points that results from selecting only those cells containing ones in the 3D occupancy grid and converting them to Cartesian coordinates.
\end{itemize}

The proposed method uses a neural network to produce the 3D occupancy grid only with radar data, using the lidar data as ground truth, as shown in the block diagram in Fig.~\ref{High-Level-Block}. As can be seen, the data collected with both sensors must be pre-processed before it can be used for the task. For the radar, signal processing has to be applied to estimate the radar cube, while for the lidar, the 3D occupancy grid has to be generated. The next subsections explain in detail both processes.

\begin{figure*}[t]
\subfloat[]{\includegraphics[width=.33\textwidth]{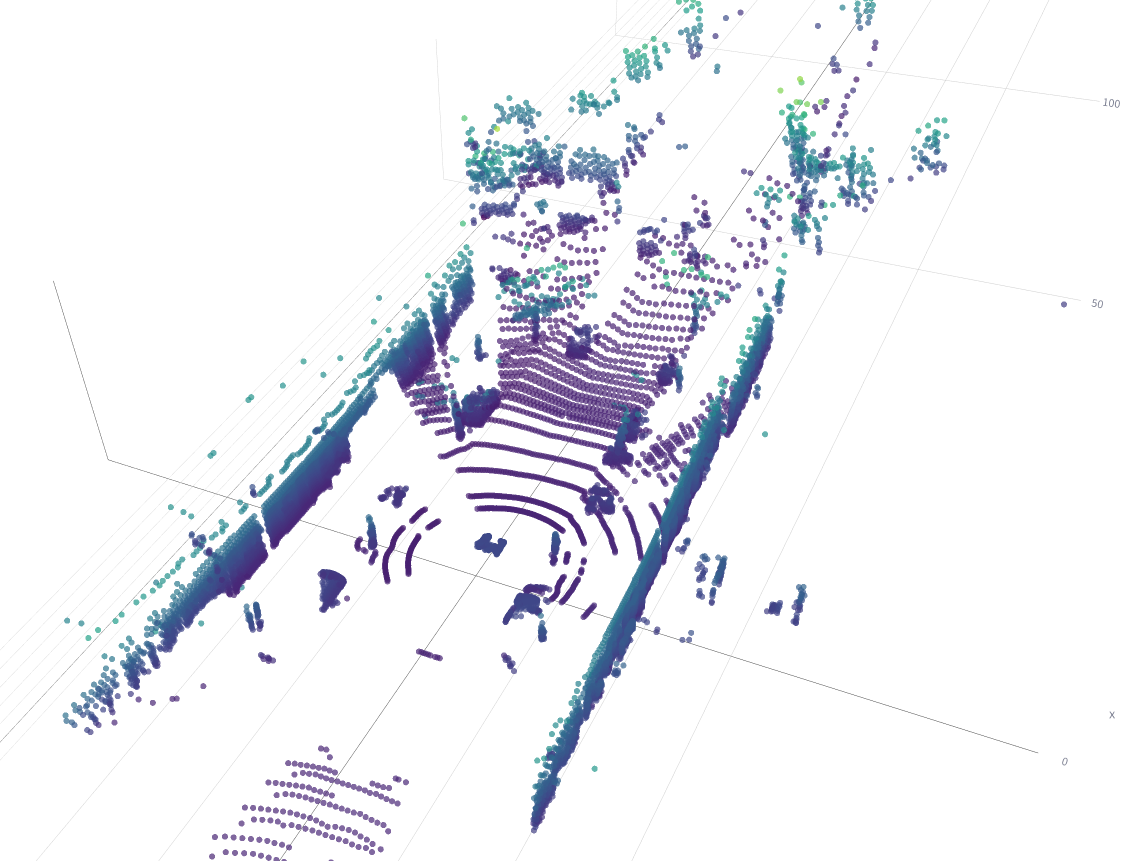}%
\label{LidarPCFull}}
\subfloat[]{\includegraphics[width=.33\textwidth]{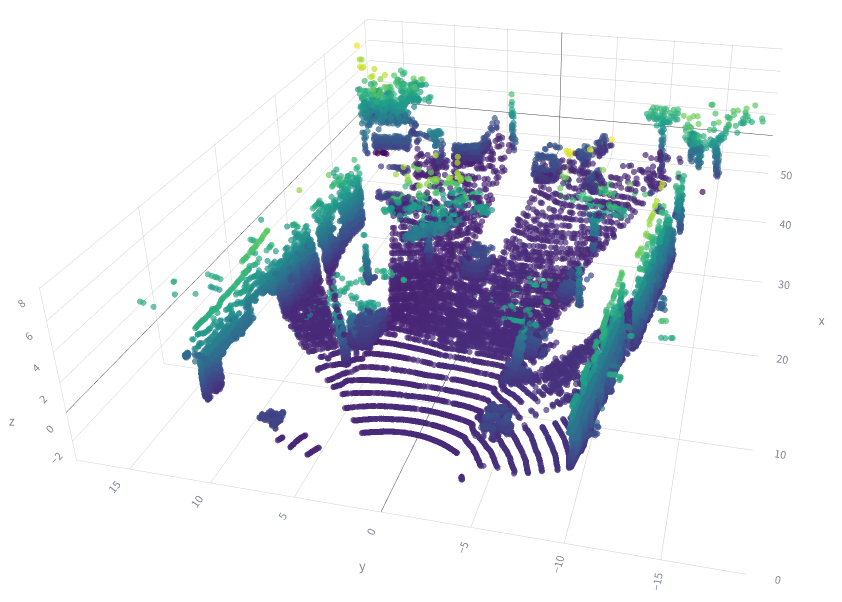}%
\label{LidarPCRoI}}
\subfloat[]{\includegraphics[width=.33\textwidth]{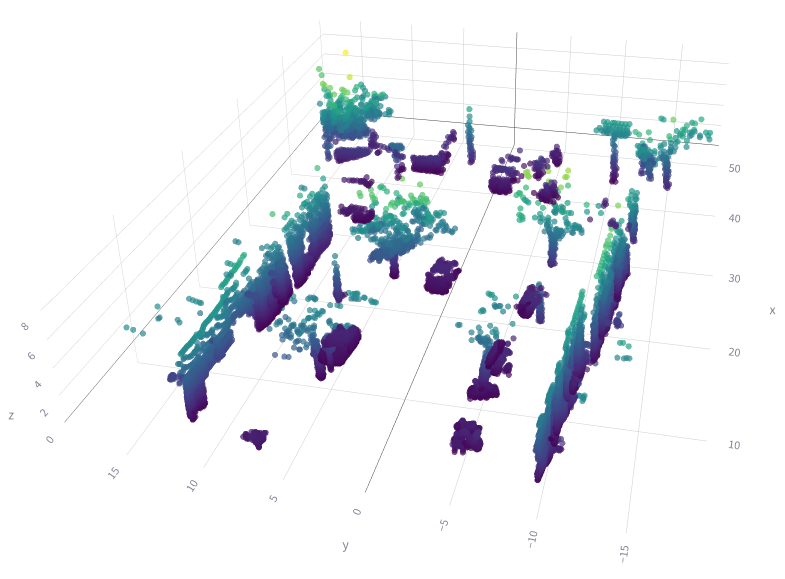}%
\label{LidarPCFinal}}
\caption{In (a) the original lidar point cloud as provided by the sensor. In (b) the lidar point cloud after the RoI cropping to mimic the radar field of view (i.e., $\pm 70^{\circ}$ in azimuth and $\pm 20^{\circ}$ in elevation). In (c), the point cloud after the road surface removal using PatchWork++ \cite{Lee2022}.}
\label{LidarPC}
\end{figure*}

\subsection{Radar Signal Processing}
\label{subsec:signa_proc}
Frequency-modulated continuous-wave (FMCW) radar with multiple-input multiple-output (MIMO) arrays has been established as the standard for automotive radars \cite{Sun2020}. While several strategies to ensure orthogonality between transmitted waves are possible \cite{Hongbo2014}, the most extended one is time division multiple access (TDMA) due to its ease of implementation. However, this introduces two undesirable effects that need to be compensated. First, the PRI (Pulse Repetition Interval) is enlarged by a factor of the number of transmitters; therefore, the maximum unambiguous Doppler is reduced, as can be seen in equation (\ref{vmax}),
\begin{equation}
    v_{max} = \frac{c}{4f_cPRI},
    \label{vmax}
\end{equation}
where $c$ is the speed of light and $f_c$ is the carrier frequency.

This effect is especially problematic in the automotive context, where targets can have high relative speeds. Moreover, the phase difference between signals received from different transmitters will depend on both the angle of arrival of the signal and the velocity of the targets (due to the target's movement between transmission times) \cite{Zoeke2015}. This phase migration term is shown in equation (\ref{phasemigration}),
\begin{equation}
    \phi_{mig} = \frac{4\pi}{\lambda}v\Delta t,
    \label{phasemigration}
\end{equation}
where $\lambda$ is the wavelength, $v$ is the relative speed of the target and $\Delta t$ is the time difference between transmitters.
This term must be compensated before performing angle estimation to avoid significant estimation distortions. Both undesirable effects of TDMA are solved using the overlapped virtual antennas present in the radar system with the algorithms provided in \cite{Schmid2012}.

An overview of the radar signal processing pipeline is shown in the bottom row of Fig.~\ref{High-Level-Block}. As it can be seen, apart from the TDMA compensation step, it is a standard processing pipeline, with the peculiarity that only a single value in elevation is saved per range-azimuth-Doppler cell. This is done for two reasons. First, due to the sparse antenna system in the vertical direction, there are grating lobes in the elevation plane. Thus, the elevation dimension is cropped in the $\pm$20 degree region to avoid the grating lobes, and the highest peak is selected as a single estimated elevation value. Second, the output data size is significantly reduced this way, from a full range-azimuth-elevation-Doppler (RAED) cube to a range-azimuth-Doppler (RAD) cube with two channels, one with the power and one with the elevation. This eases data handling for later network training. This processing pipeline can be summarized as follows:

\begin{enumerate}
  \item 2D FFT with a Hamming window in the range and Doppler dimensions.
  \item Maximum unambiguous Doppler extensions and phase migration compensation due to TDMA.
  \item Zero filling for missing virtual elements, and 2D FFT in azimuth and elevation.
  \item Region of interest (RoI) cropping in elevation at $\pm20^{\circ}$ and single peak selection.
  \item The output consists then of two range-azimuth-Doppler 3D cubes, one with the power values of each cell, and one with the elevation estimation in each cell.
\end{enumerate}

\subsection{Lidar Processing}
In order to use the lidar data as ground truth, it has to be converted into a 3D occupancy grid, as shown in Fig.~\ref{High-Level-Block}. The first step is to time synchronize the data using the radar timestamps and perform an extrinsic calibration using precomputed parameters obtained measuring reference targets. Then, the lidar data is cropped to the same field of view of the radar, i.e., $\pm 70^{\circ}$ in azimuth and $\pm 20^{\circ}$ in elevation, with a maximum range of 50m. An example of this cropped RoI is shown in  Fig.~\ref{LidarPCRoI} compared to the original in Fig.~\ref{LidarPCFull}. Moreover, removing all the returns from the road surface is essential since the road surface is hardly visible to the radars. Thus, these points would provide noisy supervision for training the network. The Patchwork++ algorithm is used to this end \cite{Lee2022}. The resulting lidar point cloud after removing the road surface points can be seen in Fig.~\ref{LidarPCFinal}. 
Finally, the lidar point cloud has to be voxelized into a cube. The voxelization process can be understood as generating a 3D occupancy grid, where each voxel contains `one' if at least one lidar point is inside and `zero' otherwise. However, it is important to note that the radar cube grid is not uniform due to the Fourier Transform processing and its relationship with the $\sin$ of the estimated angle. This effect, which makes the cells thinner at boresight and broader at the edge of the field of view, must be considered to generate the same non-uniform lidar 3D occupancy grid.


\subsection{Neural Network}
\begin{figure}
\hspace*{0.8cm}
\includegraphics[width=0.58\linewidth]{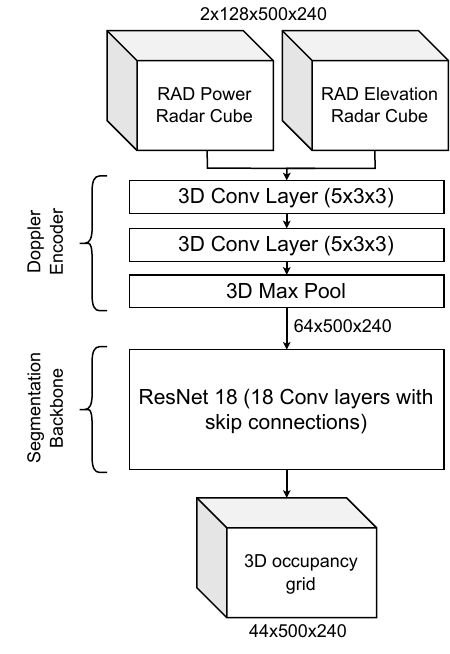}
\caption{Block diagram of the proposed neural network. The first part is a Doppler encoder network to reduce the Doppler dimension. Then, a Resnet18 segmentation backbone is used to estimate the 3D occupancy grid.}
\label{BlockDiagramNetwork}
\end{figure}

The network's inputs are two range-azimuth-Doppler radar cubes with dimensions $R\times A \times D$, where in practice $R=500, A=240$ and $D=128$. These values are higher than the number of fast-time samples, the number of chirps, and the number antenna elements due to zero padding before the FFT processing. One radar cube contains information about the power in each voxel while the other encodes the selected elevation bin, as described in Subsection \ref{subsec:signa_proc}. However, the lidar ground truth does not include Doppler information, and therefore, the Doppler dimension should be processed before estimating the 3D occupancy grid. It is known that there is a dependency between the Doppler and the angle for moving targets (or moving platforms) \cite{Yuan2023,Zhang2020}, and therefore, the Doppler information can be used to enhance the angular resolution. To this end, the first part of the network is designed to extract all the Doppler information in each range-azimuth cell and encode it into the channel dimension. This is achieved by using two 3D convolutional layers followed by a 3D max pool layer, transforming the $2 \times R \times A \times D$ input tensor into a $64\times R \times A$ tensor, where the 64 channel dimension contains the encoded information of Doppler and elevation. 
%
%
%
Then, an off-the-shelf 2D CNN backbone is applied to estimate the $R \times A \times E$ ($500 \times 240 \times 44$) 3D occupancy grid. The significant advantage of using such 2D CNN backbones is their compatibility with hardware accelerators (e.g., GPUs and TPUs) and major machine learning frameworks (e.g., TensorFlow, PyTorch), leading to enhanced computational efficiency. While our current implementation employs ResNet18 \cite{he2015deep}, our modular design allows for different backbones, enabling the system to be tailored to the specific memory and computational requirements of the intended platform.
Fig. \ref{BlockDiagramNetwork} shows a schematic representation of the proposed network. Since the data is highly imbalanced (mostly empty), the focal loss \cite{lin2018focal} between the predicted 3D occupancy grid and the ground truth lidar 3D occupancy grid is used, instead of using the standard binary cross-entropy, setting the sparsity regularizer to 0.95.

\section{Data Collection}
\label{sec:data}
To support our experiments, a new synchronized and calibrated multi-sensor dataset was recorded, similar to the View-of-Delft dataset \cite{apalffy2022} and using the same demonstrator vehicle. Specifically for this paper, a Texas Instrument MMWCAS-RF-EVM \cite{Tidep} radar board has been installed on the roof next to a RoboSense Ruby Plus Lidar (128 layer rotating lidar\footnote{https://www.robosense.ai/en/rslidar/RS-Ruby\_Plus}).  The radar board is a Multiple-Input Multiple-Output (MIMO) radar with four cascaded chips resulting in 16 receivers and 12 transmitters. The resulting virtual array is an 86-element half-wavelength spaced array in the $x$-direction (used for azimuth estimation) and a 4 Minimum Redundancy Array in the $z$-direction (elevation estimation).

\begin{table}
\caption{Radar Waveform Parameters used in the Data Collection\label{waveformParameters}}
\centering
\begin{tabular}{ @{}lr @{}}
\toprule
\textbf{Parameter} & \textbf{Value}\\
\midrule
Start Frequency (GHz) & 76\\
Effective Bandwidth (MHz) &  750\\
Chirp Slope (MHz/$\mu$s) &  35\\
Chirps Length ($\mu$s) & 28 \\
Number ADC Samples per Chirp & 256 \\
Number of Chirps per Frame& 128 \\
Sampling Frequency (Msps) & 12 \\
\bottomrule
\end{tabular}
\end{table}

The data collection took place in various real-life scenarios with different object characteristics, such as suburban, campus, and Delft old-town locations. The recordings have been organized into seven independent scenes of 5 minutes, yielding around 3000 radar frames per scene (with a frame rate of 10Hz) with their associated lidar point cloud with a total of 21000 frames. Both sensors have been time-synchronized using the provided timestamps and spatial calibrated using reference targets. The dataset has been split into training, validation, and test sets by assigning five scenes to the training and validation and two to the test. For the train validation split within the five scenes, 10\% of the frames are used for validation.
\begin{figure*}[t]
\centering
\subfloat[]{\includegraphics[width=.325\textwidth]{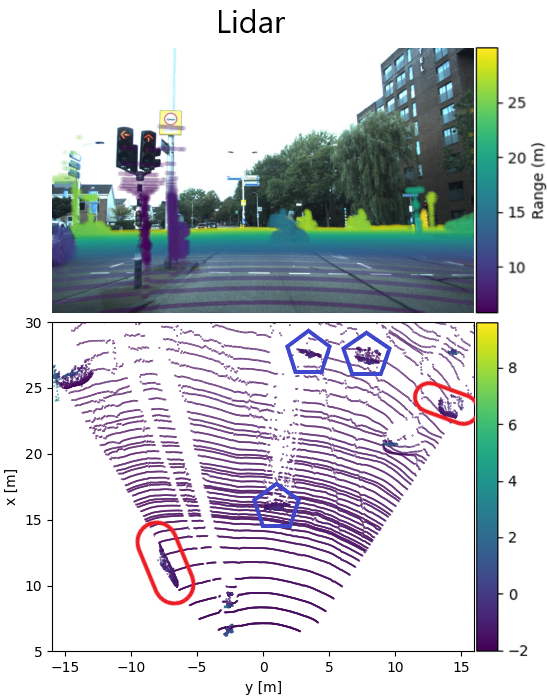}%
\label{resultsLidar}}
\subfloat[]{\includegraphics[width=.325\textwidth]{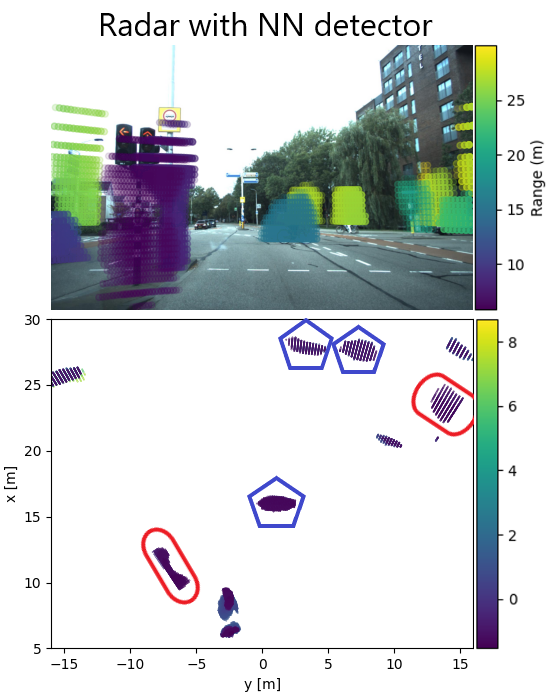}%
\label{resultsNN}}
\subfloat[]{\includegraphics[width=.335\textwidth]{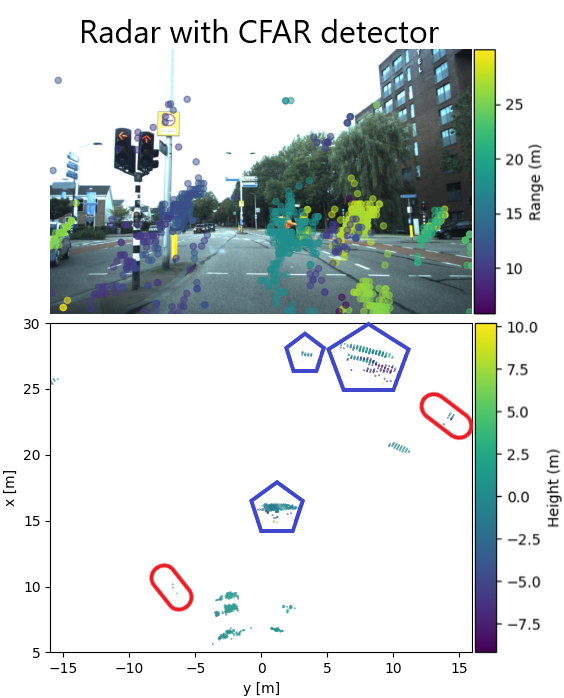}%
\label{resultsCfar}}
\caption{Data frame example \#1. In (a), the original lidar point cloud projected into the camera as well as a top-down view. In (b), the radar point cloud generated with the proposed method. In (c) the radar point cloud generated with the best implemented CFAR (2D OS-CFAR in range-azimuth followed by an OS-CFAR in Doppler).}
\label{Results}
\end{figure*}
\begin{figure*}[!h]
\centering
\subfloat[]{\includegraphics[width=.325\textwidth]{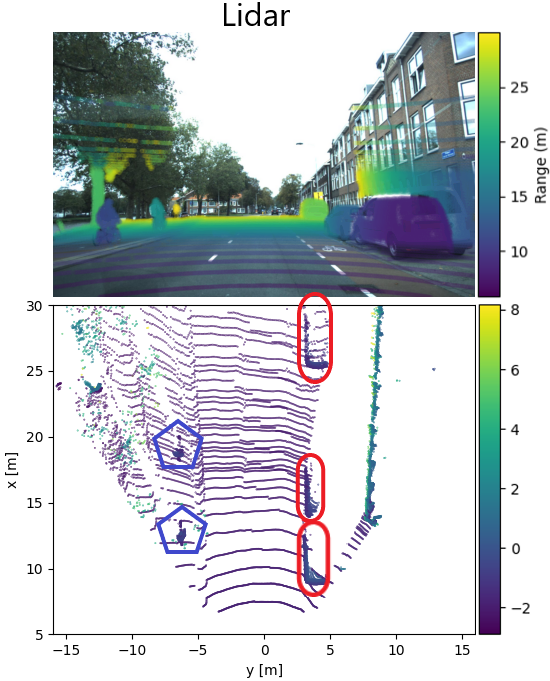}%
\label{resultsLidar2}}
\subfloat[]{\includegraphics[width=.325\textwidth]{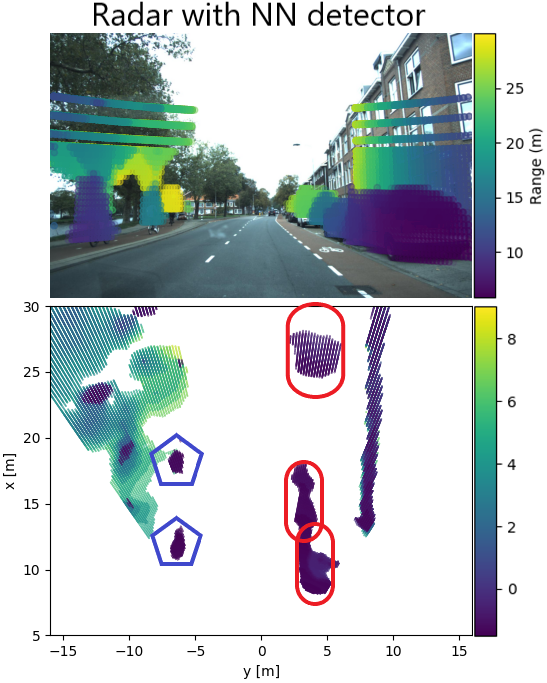}%
\label{resultsNN2}}
\subfloat[]{\includegraphics[width=.335\textwidth]{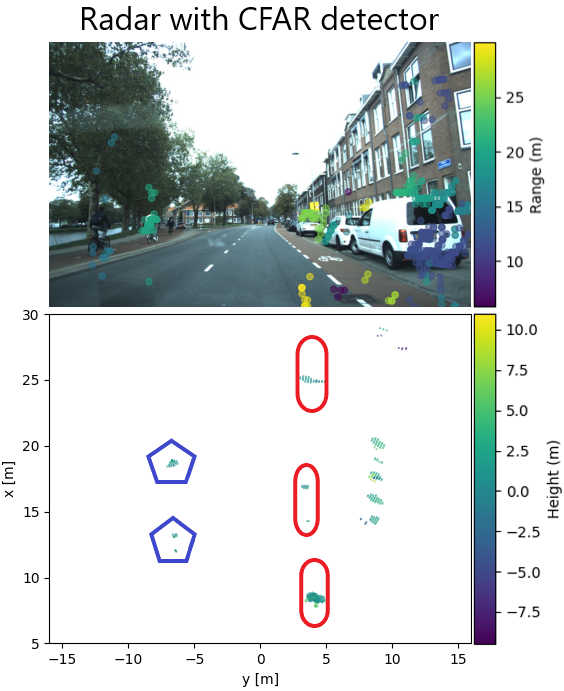}%
\label{resultsCfar2}}
\caption{Data frame example \#2. In (a), the original lidar point cloud projected into the camera as well as a top-down view. In (b), the radar point cloud generated with the proposed method. In (c) the radar point cloud generated with the best implemented CFAR (2D OS-CFAR in range-azimuth followed by an OS-CFAR in Doppler).}
\label{Results2}
\end{figure*}
\section{Results}
\label{sec:results}
After training, the proposed neural network can be used to estimate the 3D occupancy grid for each frame. The results presented in this section have been obtained using only the test set, which contains unseen data for the proposed method. The results are compared against different types of CFAR detectors with different parameters, trying always to tune them to achieve the best performance. However, optimizing CFAR parameters in every realistic driving situation is hardly possible due to the wide variety of conditions and targets. Theoretically, the order statistics (OS)-CFAR yields better results in multi-target situations \cite{Rohling83}. However, the OSCA-CFAR algorithm proposed in \cite{Kronauge2013} has also been tested since the computational complexity is lower while keeping good performance. Moreover, to avoid expensive 3D detectors, in all cases, a 2D CFAR in either range-Doppler or range-azimuth matrices has been applied, followed by a 1D CFAR or a peak detector in the remaining dimension of the radar data. For each CFAR implementation, different combinations of their hyperparameters have been tested for a fair evaluation, but only the top five best are reported in this paper.

Qualitative results of the ground truth and generated point cloud of the baseline and our proposed method are shown in Fig.~\ref{Results}. All the point clouds have been projected into the camera image to give a sense of the 3D scene (top), but the bird's eye view projection is also shown (bottom). For simplicity, the point clouds have been cropped to a maximum range of 30 meters. Moreover, as a visual aid in the top view, cyclists are highlighted with a blue hexagon, and cars are highlighted with a red ellipsoid. 
In Fig.~\ref{resultsLidar}, the original lidar point cloud is presented, where many details of the scene can be appreciated. Fig.~\ref{resultsNN} shows the detections generated using the proposed neural network, and as can be seen, most of the details are preserved. Objects are slightly overestimated in size, but the overall scene is clear. Finally, Fig.~\ref{resultsCfar} shows the output of the quantitatively best CFAR (i.e., a 2D OS-CFAR in range-azimuth followed by a 1D OS-CFAR in Doppler). 
In the top-down view, it is possible to appreciate how points for the main targets are detected, but the overall structure of the point cloud is much sparser and noisier. 
For further evaluation, a different frame is shown in Fig.~\ref{Results2} with the same approach, highlighting again the cars and cyclists.

For quantitative evaluation, two different metrics are presented. First, an analysis of the probability of detection and probability of false alarms has been performed, where the lidar 3D occupancy grid is used as the ground truth. However, these metrics do not capture any spatial relationship, i.e., a slight shift due to sensor misalignment in one of the dimensions would make the values of these two probabilities rather poor, even if the overall spatial estimation may be reasonable. For this reason, a point-cloud level metric has been included using the Chamfer distance between the lidar point cloud and the generated radar point clouds. The Chamfer distance is a common metric used to evaluate how similar two point clouds are \cite{duan20193d, hermosilla2019total, fan2016point}. For each point in each set (the sets do not need to have the same number of points), the nearest neighbor is found, and the distances are squared and summed up. The formal definition can be seen in (\ref{chamfer}),
\begin{equation}
    d(S_1,S_2) = \sum_{x\in S_1} \min_{y\in S_2} ||x-y||^2_2 + \sum_{y\in S_2} \min_{x\in S_1} ||x-y||^2_2.
    \label{chamfer}
\end{equation}

Table \ref{results}  shows the results for the top 5 best CFAR detectors and the proposed method. The CFAR guard cells in each dimension have been set to 10 in range and 16 in Doppler and azimuth. For the OS-CFAR, the rank has been set to 0.75 times the number of training cells and no guard cells have been used as recommended in \cite{Rohling83}. Only those peaks with a minimum height of 10dB below the maximum have been selected for peak detection. An important remark is needed regarding the results in Table II. As mentioned before, the computation of the $P_{fa}$ and $P_d$ is highly dependent on a good alignment of the radar cubes. A shift on a single resolution cell may be enough to reduce the $P_d$ drastically. Therefore, it is possible that a calibration misalignment exists between the lidar and the radar that the neural network is learning to resolve, significantly boosting the $P_d$ compared to the conventional CFAR approaches. Moreover, CFAR results are much more sparse than lidar data; thus, the $P_d$ is dropping. Even with this in mind, the proposed method has an excellent performance of $\sim$65\% $P_d$ compared with lidar. On the other hand, the relatively high $P_{fa}$ ($\sim$2\%) can be explained by the overestimation of most targets to the neighboring cells, which, while not desirable, may not be critical for most automotive situations. In terms of the Chamfer distance, where the possible misalignment does not play a significant role, the proposed method still outperforms all the CFAR approaches.

\begin{table}
\caption{Performance Results. RAD stands for range, azimuth, and Doppler, meaning it is the dimension where each detector is applied.\label{results}}
\centering
\begin{tabular}{@{}cccc@{}}
\toprule
\textbf{Method} & \boldmath{$P_d$ (\%)} & \boldmath{$P_{fa}$ (\%)} & \textbf{Chamfer distance ($m^2$)} \\
\toprule
Proposed Method & \textbf{65.32} & 1.99 & \textbf{1.60}\\
\midrule
\makecell{2D OS(RA) +\\1D OS(D)} & 1.72 & 0.09 & 4.98\\
\midrule
\makecell{2D OS(RD) +\\1D OS(A)} & 1.51 & 0.08 & 5.24\\
\midrule
\makecell{2D CAOS(RA) + \\1D OS(D)} & 0.22 & 0.07 & 6.98\\
\midrule
\makecell{2D CAOS(RA) + \\Peak Detector(D)} & 0.15 & 0.03 & 6.40\\
\midrule
\makecell{2D CAOS(RD) + \\Peak Detector(A)} & 0.63 & \textbf{0.02} & 4.85\\
\bottomrule
\end{tabular}
\end{table}

\section{Conclusions}
\label{sec:conclusion}
In this paper, a novel approach to radar target detection leveraging the capabilities of neural networks is presented. A comprehensive dataset comprising over 30 minutes of real-world driving scenarios has been collected using a vehicle equipped with both lidar and radar sensors, resulting in 21000 radar frames with their corresponding lidar ground truth. The dataset contains a variety of scenes such as suburban and crowded city center environments.

The work thoroughly compares the proposed neural network-based radar detector against various Constant False Alarm Rate (CFAR) methods. The results reveal that the proposed method consistently outperforms conventional CFAR techniques across various challenging scenarios. Two different types of metrics have been used: the classical probability of detection and probability of false alarm and the Chamfer distance, a point cloud level metric used to capture the spatial relationships and similarity between point clouds. The proposed neural network-based detector achieves enhanced performance in both metrics, especially in dynamic and cluttered environments. 

The success of this approach opens avenues for future research, encouraging the exploration of innovative machine-learning techniques for radar signal processing; therefore, after suitable organization, the dataset used here will be publicly released in the near future, containing the raw radar data.


\bibliographystyle{IEEEtran}
\bibliography{refs}

\begin{thebibliography}{10}
\providecommand{\url}[1]{#1}
\csname url@samestyle\endcsname
\providecommand{\newblock}{\relax}
\providecommand{\bibinfo}[2]{#2}
\providecommand{\BIBentrySTDinterwordspacing}{\spaceskip=0pt\relax}
\providecommand{\BIBentryALTinterwordstretchfactor}{4}
\providecommand{\BIBentryALTinterwordspacing}{\spaceskip=\fontdimen2\font plus
\BIBentryALTinterwordstretchfactor\fontdimen3\font minus \fontdimen4\font\relax}
\providecommand{\BIBforeignlanguage}[2]{{%
\expandafter\ifx\csname l@#1\endcsname\relax
\typeout{** WARNING: IEEEtran.bst: No hyphenation pattern has been}%
\typeout{** loaded for the language `#1'. Using the pattern for}%
\typeout{** the default language instead.}%
\else
\language=\csname l@#1\endcsname
\fi
#2}}
\providecommand{\BIBdecl}{\relax}
\BIBdecl

\bibitem{Bilik2019}
I.~Bilik, O.~Longman, S.~Villeval, and J.~Tabrikian, ``The rise of radar for autonomous vehicles: Signal processing solutions and future research directions,'' \emph{IEEE Signal Processing Magazine}, vol.~36, no.~5, pp. 20--31, 2019.

\bibitem{Sun2020}
S.~Sun, A.~P. Petropulu, and H.~V. Poor, ``{MIMO} radar for advanced driver-assistance systems and autonomous driving: Advantages and challenges,'' \emph{IEEE Signal Processing Magazine}, vol.~37, no.~4, pp. 98--117, jul 2020.

\bibitem{Palffy2020}
A.~Palffy, J.~Dong, J.~F. Kooij, and D.~M. Gavrila, ``{CNN} based road user detection using the 3d radar cube,'' \emph{IEEE Robotics and Automation Letters}, vol.~5, no.~2, pp. 1263--1270, 2020.

\bibitem{Richards2015}
M.~A. Richards, J.~A. Scheer, and W.~A. Holm, \emph{Principles of Modern Radar: Basic Principles}.\hskip 1em plus 0.5em minus 0.4em\relax Scitech Publishing Inc, 2010, vol.~1.

\bibitem{Yoon2019}
J.~Yoon, S.~Lee, S.~Lim, and S.-C. Kim, ``High-density clutter recognition and suppression for automotive radar systems,'' \emph{IEEE Access}, vol.~7, pp. 58\,368--58\,380, 2019.

\bibitem{Brodeski2019}
D.~Brodeski, I.~Bilik, and R.~Giryes, ``Deep radar detector,'' in \emph{2019 IEEE Radar Conference (RadarConf)}, 2019, pp. 1--6.

\bibitem{Cheng2022}
Y.~Cheng, J.~Su, M.~Jiang, and Y.~Liu, ``A novel radar point cloud generation method for robot environment perception,'' \emph{IEEE Transactions on Robotics}, vol.~38, no.~6, pp. 3754--3773, 2022.

\bibitem{Lin2023}
Y.~Lin, X.~Wei, Z.~Zou, and W.~Yi, ``Deep learning based target detection method for the range-azimuth-doppler cube of automotive radar,'' in \emph{2023 IEEE 26th International Conference on Intelligent Transportation Systems (ITSC)}, 2023, pp. 2868--2873.

\bibitem{Lee2022}
S.~Lee, H.~Lim, and H.~Myung, ``Patchwork++: Fast and robust ground segmentation solving partial under-segmentation using 3d point cloud,'' in \emph{2022 IEEE/RSJ International Conference on Intelligent Robots and Systems (IROS)}, 2022, pp. 13\,276--13\,283.

\bibitem{Hongbo2014}
H.~Sun, F.~Brigui, and M.~Lesturgie, ``Analysis and comparison of {MIMO} radar waveforms,'' in \emph{2014 International Radar Conference}, 2014, pp. 1--6.

\bibitem{Zoeke2015}
D.~Zoeke and A.~Ziroff, ``Phase migration effects in moving target localization using switched mimo arrays,'' in \emph{2015 European Radar Conference (EuRAD)}, 2015, pp. 85--88.

\bibitem{Schmid2012}
C.~M. Schmid, R.~Feger, C.~Pfeffer, and A.~Stelzer, ``Motion compensation and efficient array design for tdma fmcw mimo radar systems,'' in \emph{2012 6th European Conference on Antennas and Propagation (EUCAP)}, 2012, pp. 1746--1750.

\bibitem{Yuan2023}
S.~Yuan, F.~Fioranelli, and A.~Yarovoy, ``Vehicular motion-based doa estimation with a limited amount of snapshots for automotive {MIMO} radar,'' \emph{IEEE Transactions on Aerospace and Electronic Systems}, pp. 1--15, 2023.

\bibitem{Zhang2020}
W.~Zhang, P.~Wang, N.~He, and Z.~He, ``Super resolution doa based on relative motion for fmcw automotive radar,'' \emph{IEEE Transactions on Vehicular Technology}, vol.~69, no.~8, pp. 8698--8709, 2020.

\bibitem{he2015deep}
K.~He, X.~Zhang, S.~Ren, and J.~Sun, ``Deep residual learning for image recognition,'' 2015.

\bibitem{lin2018focal}
T.-Y. Lin, P.~Goyal, R.~Girshick, K.~He, and P.~Dollár, ``Focal loss for dense object detection,'' 2018.

\bibitem{apalffy2022}
A.~Palffy, E.~Pool, S.~Baratam, J.~F.~P. Kooij, and D.~M. Gavrila, ``Multi-class road user detection with 3+1d radar in the view-of-delft dataset,'' \emph{IEEE Robotics and Automation Letters}, vol.~7, no.~2, pp. 4961--4968, 2022.

\bibitem{Tidep}
\BIBentryALTinterwordspacing
T.~I. Inc. (2019) Design guide: Tidep-01012—imaging radar using cascaded mmwave sensor reference design (rev. a). [Online]. Available: \url{https://www.ti.com/lit/ug/tiduen5a/tiduen5a.pdf}
\BIBentrySTDinterwordspacing

\bibitem{Rohling83}
H.~Rohling, ``Radar {CFAR} thresholding in clutter and multiple target situations,'' \emph{IEEE Transactions on Aerospace and Electronic Systems}, vol. AES-19, no.~4, pp. 608--621, 1983.

\bibitem{Kronauge2013}
M.~Kronauge and H.~Rohling, ``Fast two-dimensional {CFAR} procedure,'' \emph{IEEE Transactions on Aerospace and Electronic Systems}, vol.~49, no.~3, pp. 1817--1823, 2013.

\bibitem{duan20193d}
C.~Duan, S.~Chen, and J.~Kovacevic, ``3d point cloud denoising via deep neural network based local surface estimation,'' 2019.

\bibitem{hermosilla2019total}
P.~Hermosilla, T.~Ritschel, and T.~Ropinski, ``Total denoising: Unsupervised learning of 3d point cloud cleaning,'' 2019.

\bibitem{fan2016point}
H.~Fan, H.~Su, and L.~Guibas, ``A point set generation network for 3d object reconstruction from a single image,'' 2016.

\end{thebibliography}

\end{document}